\documentstyle[epsf,12pt]{article}

\begin{document}
%\tableofcontents
%\chapter{INTRODUCTION}
\begin{titlepage}
\centerline{\bf{SU(2) CHIRAL SIGMA MODEL STUDY OF PHASE TRANSITION }}
\centerline{\bf{IN HYBRID STARS }}
%\maketitle
\vspace{0.5in}
\centerline { P. K. Jena$^*$ and L. P. Singh$^+$}
\date{}
\vspace{0.25in}
\centerline{Department of Physics, Utkal University, Vanivihar,}
\centerline{ Bhubaneswar-751004, India.}
%\maketitle
\vspace{1in}
%\begin{abstract}
\centerline{\bf{Abstract}}
\vspace{.1in}

  We use a modified SU(2) chiral sigma model to study nuclear matter
component and simple bag model  for quark matter constituting a neutron 
star. We also study the phase transition of  nuclear matter to quark
matter  with the  mixed phase characterized by  two conserved charges 
in the interior of highly dense neutron stars. Stable solutions of 
Tolman-Oppenheimer-Volkoff equations representing hybrid stars are 
obtained with a maximum mass of 1.67$M_{\odot}$ and radius around 8.9 km.

%\end {abstract}
\vspace{0.5in}

\vspace{0.5in}

\vspace{1.0in}
\noindent
$^*$email: pkjena@iopb.res.in \\
$^+$email: lambodar@iopb.res.in

\end{titlepage}

\section{Introduction}

  The existence of quark matter in neutron stars have been considered
by many authors$^{1,2}$ over the past two decades. In a simple but fairly 
accurate representation, a neutron star is constituted mostly of
neutrons and a small number of protons whose charge is balanced by
the  leptons. In  works prior to 1992, local charge neutrality was
imposed on all possible phase transitions above saturation density,
leaving the systems with only one conserved quantity, namely the baryon
number. Such a system is described as one component system.  
However, in 1992, Glendenning$^{3}$ pointed out that neutron star
matter has two independent conserved components, namely the baryon number and
the electric charge  and hence charge neutrality  must be applied as a
global and not a local condition. For a one component system, the
transition takes place at a constant pressure where the two phases
having different densities, for all proportions of the two phases are in
equilibrium. The equilibrium pressure is found through Maxwell
construction$^{4}$. For a two-component system, the common pressure and
all properties of the two phases in equilibrium vary with  the   
proportion  of the two phases and this region of two co-existing
phases in equilibrium is usually referred to as the mixed phase$^{3}$.

  In this paper, we have  extended our earlier one component   
analysis$^{5}$ of neutron star
structure and associated phase transition  using a two component
formalism. We consider the baryon chemical potential ($\mu_B$) and
electric charge chemical potential($\mu_E$) as the two conserved
charges.  Using Gibb's criteria for a two-component system a  first
order phase transition between beta-stable
nuclear matter and quark matter is indicated where pressure and
density vary continuously in the mixed phase region. Taking the
existence of such a phase transition between nuclear
matter and quark matter as a guide, we solve the
Tolman-Oppenheimer-Volkoff(TOV) equations with appropriate nuclear
matter$^{5}$ and quark matter$^{6}$ equations of state and find the
hybrid stars to consist of a quark-matter core with the nuclear matter 
forming a thin crust.

  This paper is organized as follows. In sec.2, we present the equation
of state(EOS) for nuclear matter. Then we proceed for  equation of state
of quark matter presented in sec.3. In sec.4, phase transition is
discussed. In sec.5, we discuss the structure of hybrid stars.  
Finally we  discuss and summarize our results in sec.6.

\section{\bf{Hadronic Phase}}

   We have used the modified SU(2) chiral sigma
model(MCH)$^{7}$ for hadronic phase since chiral model has been
very  successful  in describing high density nuclear matter. 
The importance of chiral symmetry$^{8}$ in the study of nuclear matter 
was first emphasized by Lee and Wick$^{9}$  and has become over the years  
one of most useful tools to study high density nuclear matter 
at the microscopic level.   
The nonlinear terms in the chiral sigma model give rise to the three-body 
forces which become significant in the high density regime$^{10}$. Further, 
the energy per nucleon at saturation needed the introduction of
isoscalar field$^{11}$  in addition to the scalar field of
pions$^{12}$. We also include the interaction due to isospin triplet 
$\rho$-vector meson to describe the neutron rich-matter$^{13}$.

    The modified SU(2) chiral sigma model$^{7}$ considered by us   
includes two extra higher order 
scalar field interaction terms which ensures an appropriate
incompressibility of symmetric nuclear matter at saturation
density. Further, the equation of state(EOS) derived from this model
is compatible with that inferred from recent heavy-ion collision data$^{14}$.

   The EOS for hadronic phase is calculated by using the  Lagrangian 
density$^{5}$,
\begin{eqnarray}
 L = \frac{1}{2}(\partial_{\mu} \vec{\pi}.\partial^{\mu}\vec{\pi} + 
   \partial_{\mu}\sigma  
  \partial^{\mu}\sigma )-\frac{1}{4}F_{\mu \nu} F_{\mu \nu}-
   \frac{\lambda}{4}(x^2-x_0^2)^2 
  -\frac{\lambda B}{6m^2}(x^2-x_0^2)^3 \nonumber \\ 
   -\frac{\lambda C}{8m^4}(x^2-x_0^2)^4 - g_{\sigma}\bar{\psi }(\sigma +
  i\gamma_{5}\vec{\tau} .\vec{\pi} )\psi  
 +\bar{\psi}(i \gamma_{\mu}
  \partial ^{\mu} -g_{\omega}\gamma_{\mu}\omega ^{\mu})\psi \nonumber \\   
   +\frac{1}{2}g_{\omega}^2  x^2 \omega_{\mu}\omega ^{\mu} 
  -\frac{1}{4} G_{\mu \nu}.G^{\mu \nu}+\frac{1}{2}
   m_{\rho}^{2}\vec{\rho_{\mu}}.\vec{\rho^{\mu}}
   -\frac{1}{2}g_{\rho}\bar{\psi}(\vec{\rho_{\mu}}.\vec{\tau}\gamma^{\mu})\psi 
  \end{eqnarray}
\noindent
In the above Lagrangian, $F_{\mu \nu} \equiv \partial_{\mu}\omega_{\nu} 
 - \partial_{\nu} \omega_{\mu}$, $G_ {\mu \nu} \equiv \partial_{\mu}\rho_{\nu} 
 - \partial_{\nu} \rho_{\mu}$ and $x = (\vec{\pi}^2 +\sigma ^2)^{1/2}$, 
$\psi $ is the nucleon  isospin doublet, $\vec{\pi}$ is the 
pseudoscalar-isovector pion field, $\sigma$ is the scalar field and 
$\omega_{\mu}$, is a dynamically generated isoscalar vector field, which
couples to the conserved baryonic current
$j_{\mu}=\bar{\psi}\gamma_{\mu}\psi$. $\vec{\rho_{\mu}}$ is the
isotriplet vector meson field with mass $m_{\rho}$. B and C are
constant  coefficients 
associated  with the higher order self-interactions of the scalar field .

 As obtained in our earlier work$^{5}$,  the total energy
density($\epsilon $)  and pressure(P), for the $\beta$-stable nuclear
matter is given by 

\begin{eqnarray}
   \epsilon= \frac{m^2(1-y^2)^2}{8C_{\sigma}}-\frac{B}{12C_{\omega} C_{\sigma}}
 (1-y^2)^3 +\frac{C}{16m^2C_{\omega}^2C_{\sigma}}(1-y^2)^4 \nonumber\\
+\frac{C_{\omega}n_B^2}{2y^2}+
   \frac{\gamma}{2\pi^2} \sum_{n,p,e}\int_{0}^{k_f} k^2dk
   \sqrt{k^2+{m^*}^2} + \frac{1}{2}m_{\rho}^2 (\rho_{0}^3)^2 ,
\nonumber \\ 
  P= -\frac{m^2(1-y^2)^2}{8C_{\sigma}} +\frac {B}{12C_{\omega} C_{\sigma}}
   (1-y^2)^3-\frac{C}{16m^2C_{\omega}^2C_{\sigma}}(1-y^2)^4 \nonumber\\
 +\frac{C_{\omega}n_B^2}{2y^2}+
  \frac{\gamma}{6\pi^2}\sum_{n,p,e}\int_{0}^{k_f}
   \frac{k^4dk}{\sqrt{k^2+{m^*}^2}}+ \frac{1}{2}m_{\rho}^2 (\rho_{0}^3)^2
\end{eqnarray}
\noindent
The energy per nucleon is $\frac {E}{A}=\frac {\epsilon}{n_B}$ and the
chemical potential is  $\mu=(P+\epsilon)/n_B $.  The
concentrations of neutrons, protons and electrons can be determined using
conditions of beta equilibrium and electrical charge neutrality$^{4}$ i.e \ 
$ \mu_{n} = \mu_{p}+\mu_{e}$ ; $n_p = n_e $ , 
(here $\mu_i $ is the chemical potential of the particle species
$\it{i}$). 

   The values of five parameters $C_\sigma, C_{\omega},C_{\rho} $, B and
C occurring
in the above equations are obtained by fitting with the saturation values of 
binding energy/nucleon 
(-16.3 MeV), the saturation density (0.153 fm$^{-3}$), the symmetric
energy(32 MeV), the effective(Landau) mass
(0.85M)$^{15}$, and nuclear incompressibility ($\sim $300 MeV), in accordance
with recent heavy-ion  collision data$^{13}$ are   
 $C_{\omega}$ = 1.999 fm$^2$, $C_{\sigma}$ = 6.816
fm$^2$, $C_{\rho}$ = 4.661 fm$^2$, B = -99.985 and C = -132.246.
The equation of state for the present model is found to be softer$^{5}$ 
with respect to the original  one$^{12}$

%%**************************************************************************

\section{\bf{Quark Phase }}

We  employ the bag model in our  study of the quark phase.
We consider here the quark matter EOS which includes u, d, and s quark 
degrees of freedom in addition to electrons. We have taken the
electron, up and down quark masses to be zero$^{6}$ and the strange quark
mass is taken to be 180 MeV$^{5,16}$. As shown in Ref. 5, The energy
density and total pressure of quarks including the bag constant(B) is
given by 
\begin{equation}
 P = P_e+\sum_{f} P_f -B
\end{equation}
\noindent
\begin{equation}
\epsilon = -P+ \sum_{f} \mu_f n_f. 
\end{equation}
\noindent
Where $P_e$ is the electron pressure and $P_f$ is the pressure for
quark flavor with  $\it {f}$ = u, d or s.
The chemical equilibrium condition $\mu_d$ = $\mu_s = \mu_u+\mu_e $
written$^{5}$ in terms of baryon($\mu_B$) and electric charge($\mu_E$) 
chemical potential renders P and $\epsilon$ as functions of two
independent chemical potentials $\mu_B$ and  $\mu_E$. 
Defining the number density of quark flavor $ f$ as 
 $n_f=\frac{\partial P_f}{\partial \mu_f}$, 
the baryon number density and charge density of the quark system is given by 
\begin{eqnarray}
 n_B = \frac{1}{3} \sum_{f}n_f  \nonumber \\
 \rho = \frac{1}{3}(2n_u-n_d-n_s) \nonumber 
\end{eqnarray}
\noindent
Charge neutrality of this system requires 

\centerline {$\frac{2}{3} n_u - \frac{1}{3} n_d -\frac{1}{3}n_s-n_e = 0 $}

%%************************************************************************
\section{\bf{Phase Transition}}

For the phase transition from $\beta$-stable nuclear to quark matter,
we consider the above described hadronic phase and quark phase with a
typical  value of $\alpha_s$ = 0.5, B$^{1/4}$ = 150 MeV$^{5,16}$. The 
Maxwell construction within  one component formalism was done$^{5}$,
by assuming that the two phases are separately charge
neutral keeping pressure constant with discontinuity in energy and
number density over the transition region. The transition was found
to extend over the density range of  0.27 -  0.37 fm$^{-3}$.

  In the present work we employ theoretically more accurate     
two-component analysis  which makes
the pressure and density vary continuously over the mixed phase
region. We follow Glendenning procedure$^{3}$ of parameterising nuclear
matter by two chemical potentials ($\mu_B,\mu_E$) i.e baryon and  
electric charge chemical potentials. Gibb's condition for  phase
equilibrium at zero temperature for both phases gives 
\begin{equation}
  P_{HP}(\mu_B,\mu_E)=P_{QP}(\mu_B,\mu_E)
\end{equation}
\noindent
As described in Ref. 17, the nuclear and quark matter share a common
electron distribution in the mixture is more appropriate. Hence the
overall charge neutrality requirements is that the nuclear, quark and
lepton components together are charge neutral, i.e 
\begin{eqnarray}
\chi\rho_{QP}(\mu_B,\mu_E)+(1-\chi)\rho_{HP}(\mu_B,\mu_E)
         =\rho_e(\mu_B,\mu_E)
\end{eqnarray}
\noindent
Where $\chi $ is the filling fraction of quark matter in the mixed phase. 
The above two equations(5,6) involve three variables $\mu_B, \mu_E$ and
$\chi$. So for a given value of $\mu_B$, the two equations can be
solved simultaneously to  give the values of $\mu_E$ and $\chi $. Then 
the energy density $\epsilon_{MP}$ and baryon density $n_{MP}$ of the
mixed phase can be calculated using the equations$^{18}$
\begin{eqnarray}
\epsilon_{MP}=\chi \epsilon_{QP} + (1-\chi)\epsilon_{HP} \nonumber \\
 n_{MP}=\chi n_{QP} + (1-\chi)n_{HP}
\end{eqnarray}
\noindent
This procedure is carried out provided that $ 0 \leq \chi \leq 1 $. 
For $\chi=0$ the system is in pure hadronic  phase and for $\chi=1$ the
system is in pure quark phase. 
\begin{figure}[t]
\leavevmode
%\centerline{\psfxsize=2.5in\psfbox{pmu3.ps}}
\protect\centerline{\epsfxsize=5in\epsfysize=4.5in\epsfbox{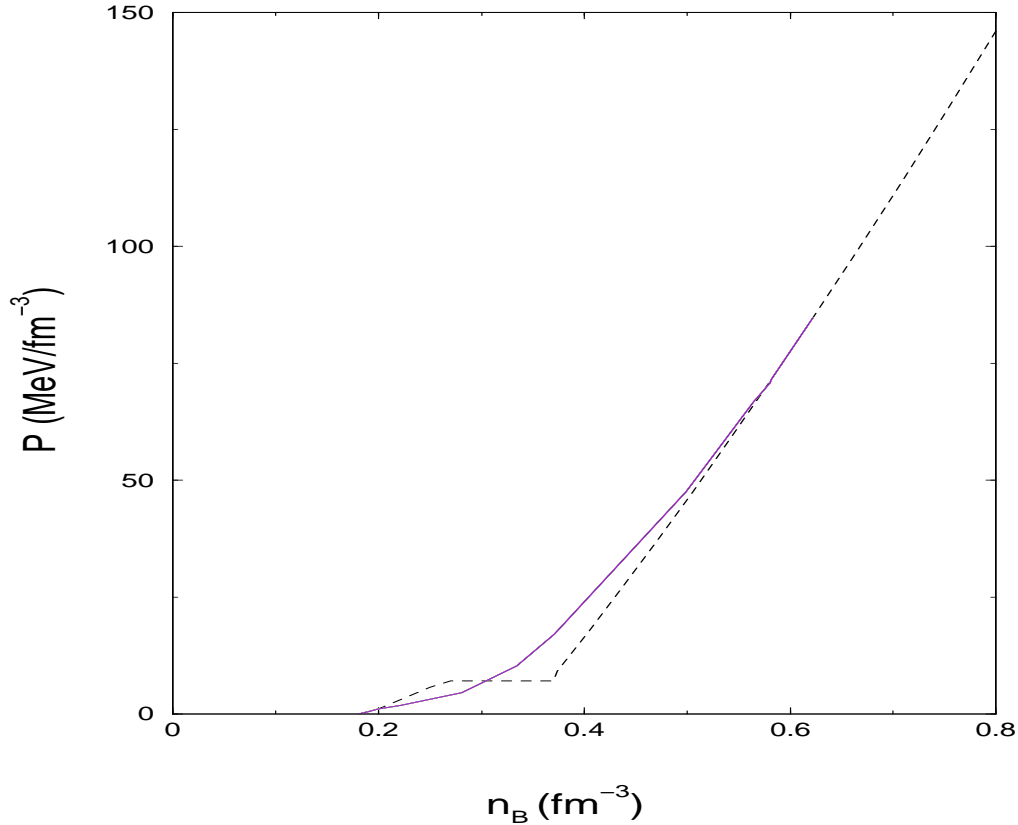}}
%\centerline{\epsfxsize=4in\epsfbox{pmu3.eps}}
\caption{\it{Pressure (P) as function of  baryon  number density
 in the mixed phase region. The solid curve corresponds to
 two-component system and the dotted for one component system.}} 
\end{figure}

  For $\alpha_s$ = 0.5, B = (150 MeV)$^4$, it is found that the number 
density and energy density of the mixed phase extends over the range
0.20 - 0.58 fm$^{-3}$ and 191 - 638 MeV/fm$^{-3}$ respectively.
The total pressure thus obtained is now no longer constant over the density
range of the mixture; it increases   monotonically in the mixed phase
region as shown in Fig.1 by solid line. In the current picture of bulk 
phases, the local charge neutrality  of both phases is relaxed to a
global charge neutrality condition which indicates that in the mixed
phase region the nuclear matter phase will be positively charged
whereas the quark matter phase will be negatively charged$^{3}$.
As the phase transition implies that the neutron star consists of
quark matter in the interior and nuclear matter in the periphery hence 
it can be called as a hybrid star.

\section{\bf{Structure of Hybrid Stars.}}

    As a hybrid star is assumed to be a spherically symmetric
distribution of mass in hydrostatic equilibrium, the equilibrium
configurations  are obtained by solving the Tolman-Oppenheimer-Volkoff 
(TOV) equations$^{19}$ for the pressure $P$ and the enclosed mass $m$,
\begin{equation}
 \frac{dP(r)}{dr} = -G \frac {[\epsilon(r)+ P(r)][m(r)+ 4\pi r^3 P(r)]}
    {r^2[1-2G m(r)/r]} \ ,  
\end{equation}

\begin{equation}
 \frac{dm(r)}{dr} = 4\pi r^2 \epsilon(r)  
\end{equation} 
\noindent
Where G is the gravitational constant, $m(r)$ is the mass contained in a 
volume of radius $r$. In order to get stellar radius $R$ and the
gravitational mass $M = m(R)$ we integrate the TOV equations for a given 
central energy density $\epsilon(r=0) = \epsilon_c$ with the
boundary condition $m(r=0) = 0$ and $P(r = R) \sim 0$.

       We have plotted in Fig.2 the mass  of hybrid star as a
function of  central energy density to examine the stability of such a 
star. Taking into account the stability of such  stars
under density fluctuations require $ dM/d\epsilon_c > 0$$^{20}$. 
As may be seen from the figure, $dM/d\epsilon_c$ becomes negative
around 1800 MeV/fm$^3$ after which it  may
collapse into black holes$^{19,20}$. This yields the maximum mass of
hybrid star as $M \simeq 1.67 M_{\odot}$. Fig.3  shows the mass as a 
function of radius obtained for different central densities  for 
such a star which indicates the maximum radius to be around 8.9 km.
\begin{figure}[h]
\leavevmode
\protect\centerline{\epsfxsize=5in\epsfysize=5in\epsfbox{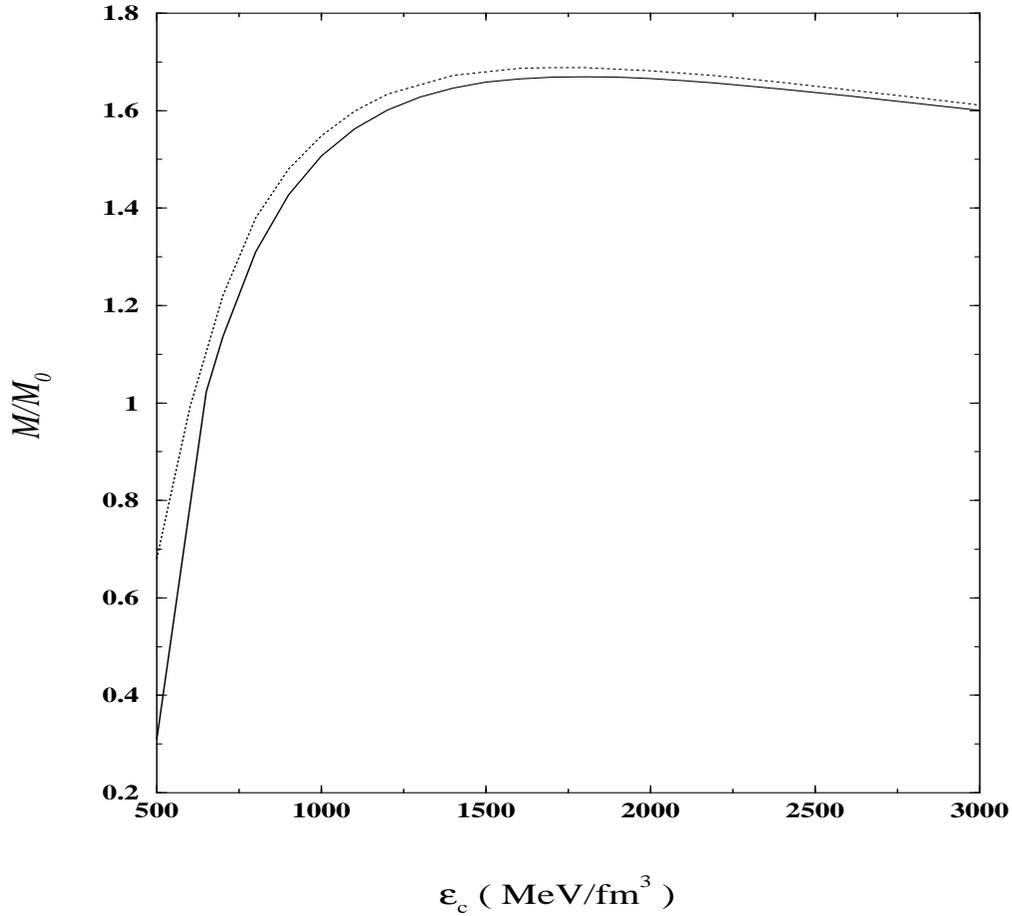}}
%\centerline{\epsfxsize=4in\epsfbox{pmu4.eps}}
\caption{\it{The mass(M/$M_{\odot}$) of the  hybrid
   star as a function of central energy density($\epsilon_c$).The
   solid curve corresponds to two-component system and the dotted for 
   one component system.}} 
\end{figure}

   We observe that for a given central energy density of 700
MeV/fm$^3$  such a star has pure
quark matter core of radius 3.29 km, mixed phase covers 5.08 km and
pure nuclear  matter crust of 0.21 km, whereas for $\epsilon_c$ = 1800 
MeV/fm$^3$, the corresponding datas are 6.69 km, 1.63 km and 0.1 km.
Hence it is clear that if we take smaller central energy density then
pure quark phase is found to cover less region and the mixed phase
covers more region out of the total radius of the hybrid star and the
reverse happens in case of larger central energy density. But pure
nuclear matter covers a thin crust of the hybrid star in the whole
range of relevant central energy densities. 
Thus the existence of a broad mixed phase and a much thinner layer of  pure
nuclear matter phase($\sim \frac{1}{5}$th  of the one-component system) are
the two most important features distinguishing the characteristics of a 
two-component system vis-a-vis a one-component one. We have exhibited the  
results of both analyses in all our figures for quick visual comparision.

    Since the data with regard to symmetric energy(32$\pm$6 MeV) and 
nuclear incompressibility (300$\pm$50 MeV) which have been used to fix 
the coupling constants of the theory are associated with some
errors, we have also tried to check the sensitivity of our results for 
$M_\odot$ and R with respect to these errors. To do so  we have computed the 
coupling constants ($C_{\rho},C_{\omega},C_{\sigma}$,B and C) for four cases 
of symmetric energy value 32 MeV with incompressibility 250 and 350 MeV and
the incompressibility of 300 MeV with symmetric values 26 and 38 MeV. The 
maximum values for M and R are found to be M = 1.67$^{+0.007}_{-0.001} 
M_{\odot}$ and R = 8.9$^{+0.29}_{-0.03}$ kms. Such small effect of
errors in  symmetric  energy and incompressibility on maximum values
of  M and R can be attributed to the fact that only about 2$\%$ of 
the radius of the hybrid star contains pure nuclear matter.    
\begin{figure}[h]
\leavevmode
\protect\centerline{\epsfxsize= 5in\epsfysize=5in\epsfbox{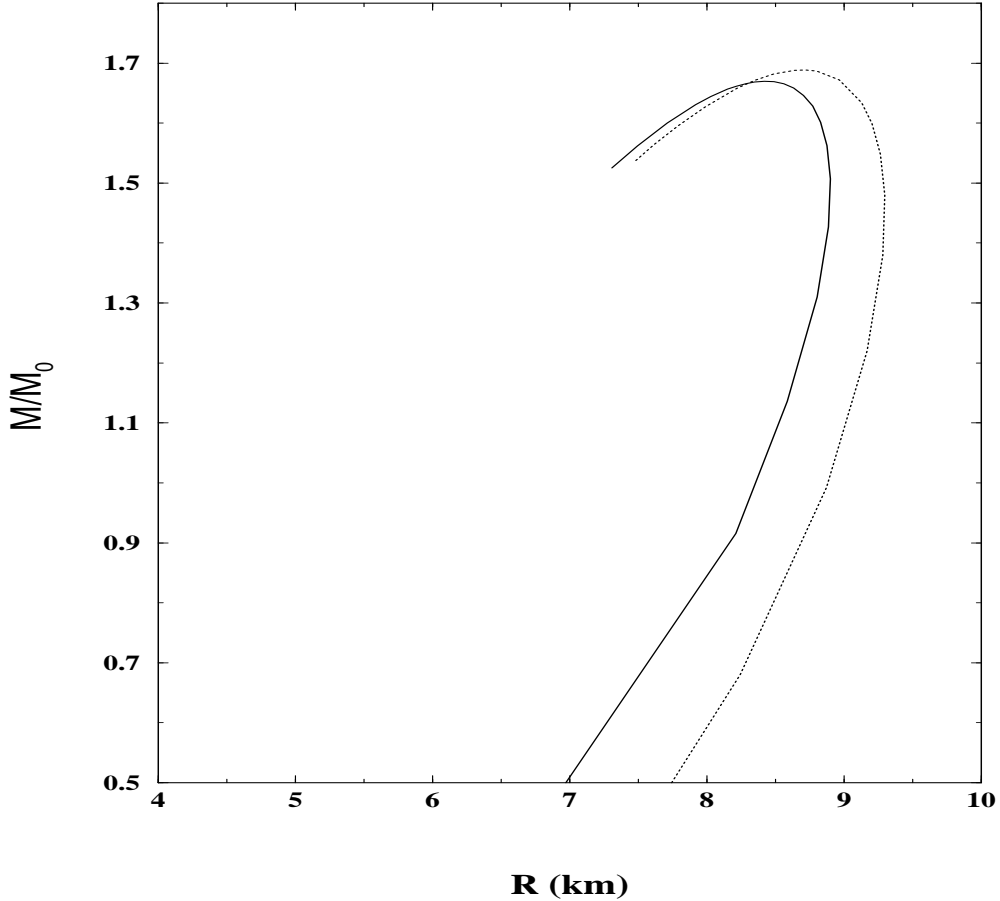}}
%\centerline{\epsfxsize=4in\epsfbox{pmu4.eps}}
\caption{\it{Mass as a function of radius for the hybrid star.
  The solid curve corresponds to two-component system and the dotted 
  for one component system.}} 
\end{figure}

      We also calculate the surface gravitational redshift $Z_s$ of photons 
which is given by$^{21}$
\begin{equation}
    Z_s = \frac{1}{\sqrt{(1-2GM/R)}}-1 
\end{equation}
\noindent
 In Fig.4, we have plotted $Z_s$ as a function of $M/M_{\odot}$ . 
In this context it may be mentioned here that  our result for the
surface redshifts lying in the range of 0.2 to 0.5 agrees quite well
with the values  determined from gamma ray bursters$^{22}$.
We then compute the relativistic Keplerian angular velocity $\Omega_k$ 
given by$^{23}$
\begin{equation}
   \frac{\Omega_k}{10^4 sec^{-1}} = 0.72 \sqrt{\frac{M/M_{\odot}}
        {(R/10km)^3}}
\end{equation}
\noindent
             
\begin{figure}[h]
\leavevmode
\protect\centerline{\epsfxsize=5in\epsfysize=5in\epsfbox{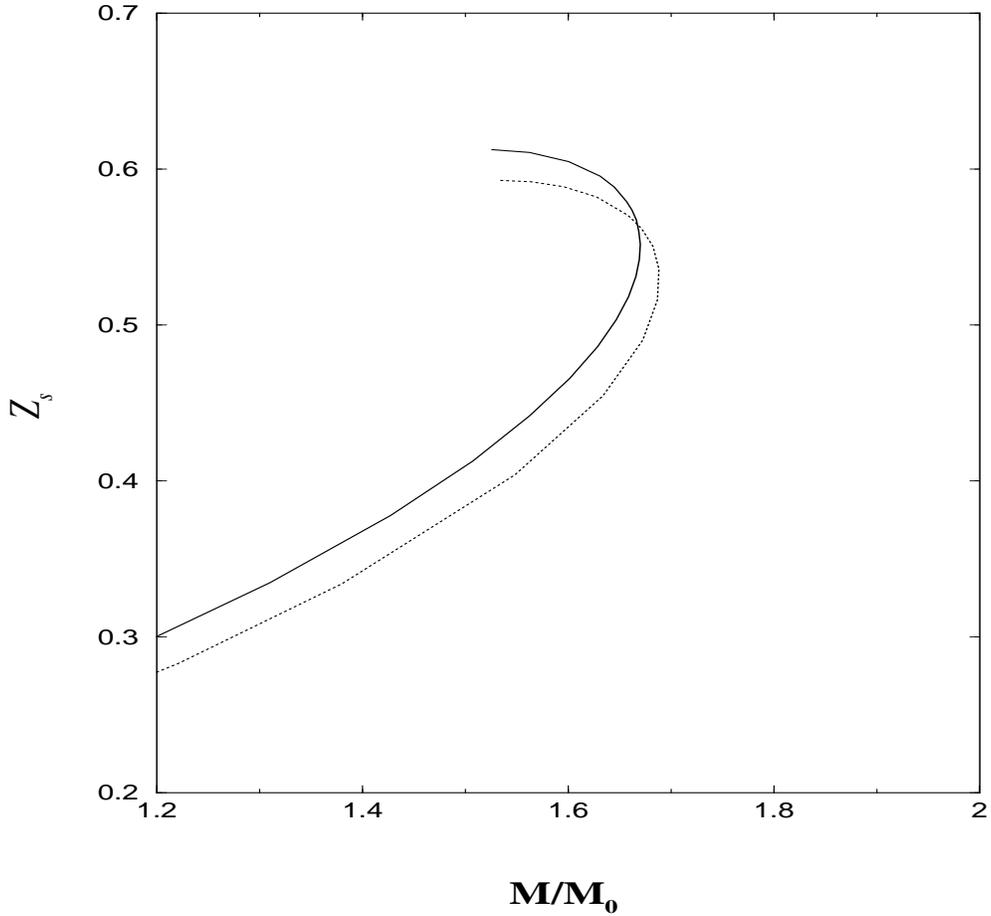}}
%\centerline{\epsfxsize=4in\epsfbox{pmu4.eps}}
\caption{\it{The surface gravitational redshift($Z_s$) as a function of star 
  mass(M/$M_{\odot}$).The solid curve corresponds to
 two-component system and the dotted for one component system.}} 
\end{figure}

\begin{figure}[h]
\leavevmode
\protect\centerline{\epsfxsize=5in\epsfysize=5in\epsfbox{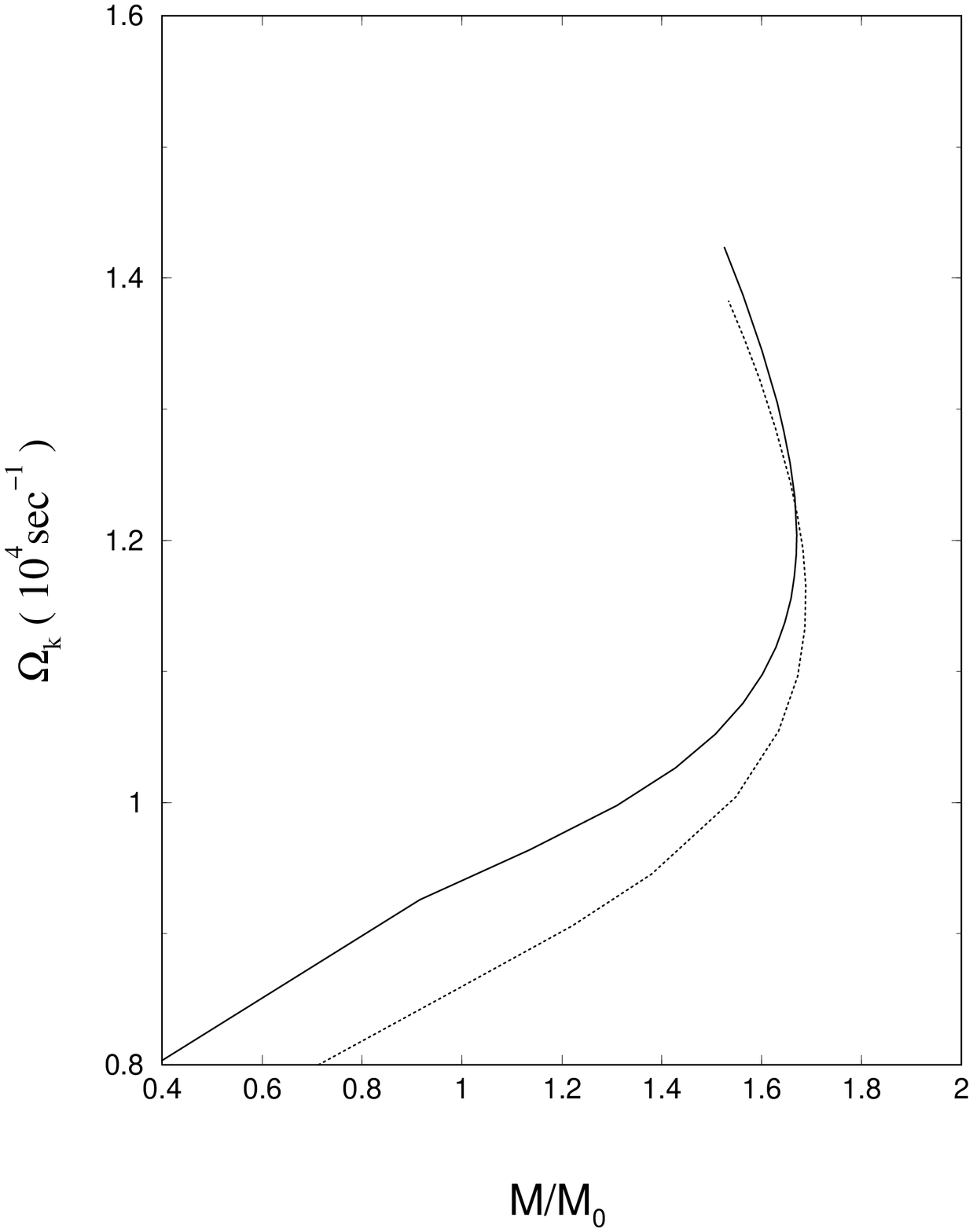}}
%\centerline{\epsfxsize=4in\epsfbox{pmu4.eps}}
\caption{\it{The Keplerian angular velocity$(\Omega_k)$ as a function of star 
  mass($M/M_{\odot}$).The solid curve corresponds to
 two-component system and the dotted for one component system.}} 
\end{figure}
\noindent
as for neutron stars. Fig.5 shows our result for variation of relativistic 
Keplerian angular velocity as a function of $M/M_{\odot}$ for such a star.
We find that the $\Omega_k$ has an inverse relationship after an
initial increase with the mass of the star beyond 1.67$M_{\odot}$.
This indicates that we can not have mass of hybrid star more than  
about 1.67$M_{\odot}$. We also observe that the maximum  value of
$\Omega_k$ being near about 1.18 $\times 10^4 $ sec$^{-1}$ implies the
maximum  period to be  0.53 millisecond that agrees with the results
obtained by Burgio et.al$^{24}$.

%%************************************************************************

\section{\bf{Conclusions}}

In this work we have extended our earlier one-component analysis$^{5}$ 
of phase transition  and structure  associated with the  hybrid stars  using 
a two-component formalism. As expected, the results of both analysis share 
similar qualitative features with certain quantitative differences.
We find that  a first order phase transition
exists between the hadronic phase and quark phase at density of about
four times the nuclear matter density. The phase transition from
nuclear matter to quark matter indicates that the core of a neutron
star consists of quark matter. To obtain stable hybrid star solution,
we have solved TOV equations using appropriate equations of state and
have taken $\alpha_s=0.5$ and B=(150 MeV)$^4$ for quark matter EOS. We 
observe that a stable hybrid star with quark  core and a nuclear 
crust exists upto $\epsilon_c$ = 1800 MeV/fm$^3$ beyond which
instability is indicated. For the values of the parameters used in 
this model we found that the maximum mass and radius of hybrid star to 
be about 1.67$^{+0.007}_{-0.001} M_{\odot}$ and 8.9$^{+0.29}_{-0.03}$ km  
respectively. It is also observed
that if we take smaller central energy density then
pure quark phase is expected to cover less region and the mixed phase
covers more region out of the total radius of the hybrid star and the
reverse happens in case of larger central energy density. But pure
nuclear matter covers a thin crust of the hybrid star in all situations.

  Using this modified SU(2) chiral model the  maximum mass and radius
of neutron star was observed in Ref.7 as M = 2.1$ M_\odot$, R = 12.1 km.
The corresponding results for hybrid star characterized by a mixed
phase  with one conserved charge obtained in our earlier work$^{5}$ was 
M = 1.69$^{+0.005}_{-0.001} M_\odot$, R = 9.3$^{+0.21}_{-0.05}$ km. 
Thus we find that  a hybrid star characterized by a mixed phase  
is more compact than a neutron star. Further we notice that sharpening 
of the analysis of the structure of hybrid star with inclusion of two
conserved charges instead of one, stabilises the star in a slightly
more compact form(M = 1.67$^{+0.007}_{-0.001} M_{\odot}$,  
R =  8.9$^{+0.29}_{-0.03}$ km). We also find that two-component analysis 
leads to  much larger mixed phase region inside the hybrid star with
corresponding shrinkage of the pure nuclear matter crust. These we
consider as important distinguishing features of the results obtained
with two and one component formalism.  Overall it is
observed  that there is a decrease in maximum mass  due to the
existence of mixed phase in the core of a hybrid star.
The greater compactness of the star also leads to smaller time period
lying in the sub-millisecond range as obtained by us. It is also
observed that the surface gravitational redshift  and relativistic
Keplerian angular velocity of the  hybrid star can not increase beyond 
$M/M_\odot$ = 1.67; showing a decrease with increase in $M/M_\odot$
beyond this value.
  
%%**************************************************************************
\vspace {0.1in}
\noindent {\bf{Acknowledgements}}
\vspace {0.05in}

    We are thankful to Institute of Physics, Bhubaneswar,
India, for providing the library and computational facility. P. K. Jena would
like to thank Council of Scientific and Industrial Research, Government of
India, for the award of SRF, F.No. 9/173 (101)/2000/EMR-I.
We thank the referee for suggesting valuable improvements in the manuscript.

%\newpage
%\section{\bf {References}}

\end{document}